 \documentclass[cameraready]{Interspeech}

\usepackage{graphicx}
\usepackage{amsmath,amssymb}
\usepackage{booktabs}
\usepackage{array}
\usepackage{tabularx}
\usepackage{siunitx}
\usepackage{url}
\usepackage[table]{xcolor}
\newcolumntype{C}[1]{>{\centering\arraybackslash}p{#1}}
\newcolumntype{R}[1]{>{\raggedleft\arraybackslash}p{#1}}

\title{Efficiency-Performance Trade-offs in Neural Speaker Diarization via Structured Pruning and Low-Bit Quantization}

  \author[correspondingauthor]{Rishit}{Chatterjee}
  \author{Tahiya}{Chowdhury}
  \address{
    Department of Computer Science, Colby College, Waterville, Maine, United States
  }
  \email{rchatt28@colby.edu , tahiya.chowdhury@colby.edu}


\begin{document}

\maketitle

\begin{abstract}
Streaming speaker diarization is crucial for time-critical medical dispatch, but deploying it on resource-constrained hardware requires smaller, faster models. Using SIMSAMU, a dataset of simulated medical-dispatch conversations, we evaluate streaming behavior before compressing the segmentation model with pruning and low-bit quantization. We characterize performance across a range of streaming latency budgets and find that additional buffering is not consistently beneficial, while very low-latency operating points can substantially degrade performance. Our study shows that model compression trades performance for memory footprint, and we highlight an operating point where FP16 reduces model size by half with essentially unchanged real-time factor, at a cost of a 40\% relative DER increase against the baseline. This work characterizes the trade-offs for real-time deployment and contributes to speech technology that can enable reliable human communication in time-critical contexts.

\end{abstract}

\section{Introduction}

Speaker diarization determines \emph{who spoke when} in multi-speaker recordings \cite{oshaughnessy25_diarization_review} and is a key upstream component for speaker-attributed automatic speech recognition (ASR), meeting transcription, and conversational analytics \cite{park21_diarization_review, anguera12_taslp}. However, reliability varies sharply across domains (e.g., meetings, broadcasts, call-center audio) due to acoustics, interaction patterns, and domain mismatch \cite{ryant21_interspeech,medennikov25_interspeech}. This is particularly consequential in safety-critical settings such as emergency medical dispatch, where accurate speaker attribution supports downstream processing in a domain- and language-specific context \cite{nun25_simsamu}. Efficiency is therefore a practical limitation: streaming systems must produce speaker labels with low latency, and batch systems must operate faster than real time within predictable memory budgets \cite{medennikov25_interspeech, rahou24_interspeech}. These constraints are amplified in modern neural diarization, where accuracy gains often rely on scaling transformer encoders, increasing parameters and inference-time memory \cite{baevski20_wav2vec2,han25_pruning_ssl_diarization}, and in operational audio with channel mismatch and background noise \cite{nun25_simsamu}. Finally, because clinical recordings can contain protected health information and speech can act as a biometric voiceprint, public release of domain-specific diarization data is uncommon, limiting systematic evidence about diarization behavior in this setting \cite{hhs_hipaa_deid,wiepert24_jmir}.

\section{Background}
\label{sec:lit}
\textbf{Efficiency-aware streaming diarization.}
Diarization is often deployed either online or at scale (requiring high throughput and predictable memory) \cite{durmus25_interspeech,kynych24_lightweight_realtime_sd}. 
A direction is to design models and training strategies for bounded-delay inference. This includes causal segmentation models with configurable look-ahead and streaming diarization models that maintain compact speaker states (e.g., buffers/caches) to preserve consistent speaker mapping under small delay budgets \cite{rahou24_interspeech, xue21d_interspeech, coria21_asru}. 
Several studies explicitly sweep look-ahead and buffering choices to quantify latency--accuracy trade-offs under controlled online inference \cite{rahou24_interspeech,xue21d_interspeech,coria21_asru,xue21_slt_stb}. Together, these studies motivate treating latency constraints and runtime budgets as explicit experimental factors when evaluating diarization systems, rather than as post hoc deployment constraints. \textit{However, few works provide an end-to-end characterization of how diarization accuracy shifts across practical streaming latency budgets under a fixed, reproducible pipeline in a medical conversational setting.}

\textbf{Pruning for resource-constrained diarization.}
Compression of self-supervised learning (SSL) speech encoders has been studied in recent speech research because modern diarization pipelines increasingly use large pretrained SSL backbones (e.g., HuBERT/WavLM-style encoders) as their acoustic feature extractors \cite{hsu21_hubert, chen22_wavlm, yi25_ell2}. However, these models are compute- and memory-intensive, motivating pruning. Prior work explores both \emph{structured} pruning of SSL transformers (e.g., removing layers/heads/feed-forward network components) and \emph{unstructured} one-shot pruning to induce sparsity \cite{han25_interspeech,peng23c_interspeech,gu24_interspeech}. \textit{While the results suggest that pruning can be a practical approach for meeting compute budgets, its impact on diarization inference under throughput and memory constraints remains underexplored.} 

\textbf{Low-bit quantization for speech models.}
Quantization is another compression method that reduces memory footprint and can accelerate inference on integer-optimized hardware. The general foundation for integer-arithmetic-only inference is well established in previous works\cite{Jacob_2018_CVPR, krishnamoorthi18_quant_whitepaper, kim21d_ibert}, with techniques such as low-bit post-training~\cite{gu25b_interspeech}, 8-bit integer~\cite{wagner24_interspeech,hong25_icassp}, and mixed precision~\cite{kang25_genptq,fish23_interspeech} found useful. 
However, fewer works have evaluated quantization's impact in end-to-end diarization pipelines. This matters because diarization combines frame-level segmentation and speaker-embedding stages, and low-precision arithmetic can affect these components differently than token-level recognition, motivating diarization-specific quantization studies\cite{aperdannier24_diart_opt,gu25b_interspeech,wagner24_interspeech, bredin23_interspeech}. \textit{As a result, the accuracy--footprint--throughput trade-offs of low-bit quantization remain less documented for diarization problem than for ASR, especially under realistic streaming constraints.}


\subsection{Contributions}
In this work, we address the aforementioned gaps, and our work makes three contributions:
\begin{itemize}{
  \item We characterize diarization accuracy across a wide range of streaming latency budgets, and show that additional buffering is not consistently beneficial, while very low-latency operating points can substantially degrade performance.
  \item  We 
  show that choice of pruning structure and level substantially influence performance, motivating careful selection of the pruning strategy.
  \item We show a case where pruning followed by quantization reduced the segmentation model size to half the original size—at a cost of a $\sim$40\% increase in error when compared to the baseline model. 
  }
\end{itemize}

\section{Data}
Our experiments use \textbf{SIMSAMU}, a public corpus of \emph{realistic medical dispatch phone dialogues} in French, annotated for diarization and transcription \cite{nun25_simsamu,hf_simsamu_dataset}. The corpus contains call recordings spanning a wide range of emergency scenarios. \cite{nun25_simsamu,llm4all_wp4_simsamu}. This domain is critical for speech technology because dispatch conversations are time-critical, highly structured, and acoustically challenging, yet public diarization resources are uncommon due to privacy constraints \cite{nun25_simsamu}. SIMSAMU is well suited to our efficiency-first questions because it provides a public, standardized diarization recipe and scoring protocol, enabling controlled streaming and compression sweeps under a fixed pipeline \cite{nun25_simsamu,hf_simsamu_dataset}. As a service-domain corpus, it also helps address the lack of end-to-end evidence for diarization under practical deployment constraints in privacy-sensitive settings. For a detailed corpus breakdown and collection protocol, we refer the reader to the SIMSAMU dataset paper \cite{nun25_simsamu}.

\section{Methods}
\subsection{Experimental Setup}
\label{sec:setup}
Our experiments follow the diarization setup provided with the SIMSAMU dataset release, i.e., we keep the same preprocessing, pipeline structure, and scoring protocol as the dataset paper and accompanying reference implementation \cite{nun25_simsamu,medkit_audio_metrics}. \footnote{We adopt the SIMSAMU pipeline (\texttt{medkit/simsamu-diarization}): a \texttt{PyanNet} speaker-segmentation model initialized from \texttt{pyannote/segmentation-3.0} and paired with a pretrained speaker-embedding extractor (\texttt{pyannote/wespeaker-voxceleb-resnet34-LM}) and agglomerative clustering \cite{hf_simsamu_diarization,hf_pyannote_segmentation30,hf_pyannote_wespeaker_resnet34lm}. The segmentation model uses a SincNet front-end followed by a 4-layer bidirectional LSTM (128 hidden units) and a 2-layer post-LSTM linear stack (128 hidden units) \cite{hf_simsamu_segmentation}.} We keep hyperparameters fixed to the defaults, including segmentation post-processing, embedding extraction options, and clustering settings, applying pruning/quantization only to the segmentation model \cite{hf_simsamu_diarization}. We keep the same evaluation protocol and report diarization error rate (DER) as the evaluation metric. All experiments were run on a single NVIDIA L40S GPU with 28 CPU cores and 64GB RAM, using FP32 inference unless stated otherwise.

\subsection{Evaluation Metric}

\textbf{DER.} DER is the standard performance metric for speaker diarization 
\cite{anguera12_taslp,ryant21_interspeech}.
DER measures the fraction of \emph{reference speaker time} that is not correctly attributed, decomposed into false alarm, missed speech, and speaker confusion \cite{dihard2_evalplan,pyannote_metrics_ref}:
\begin{equation}
\mathrm{DER} = \frac{T_{\mathrm{FA}} + T_{\mathrm{MISS}} + T_{\mathrm{CONF}}}{T_{\mathrm{REF}}}.
\end{equation}
Here, $T_{\mathrm{REF}}$ is the total reference speaker time, 
$T_{\mathrm{FA}}$ is hypothesized speaker time not attributed to any reference speaker,
$T_{\mathrm{MISS}}$ is reference speaker time not attributed to any hypothesized speaker, and
$T_{\mathrm{CONF}}$ is reference speaker time attributed to the wrong speaker \cite{dihard2_evalplan,pyannote_metrics_ref}.

DER is computed after finding an optimal one-to-one mapping between reference speakers and hypothesized speaker labels (using Hungarian algorithm), so that errors reflect speaker attribution under the best label alignment \cite{dihard2_evalplan,pyannote_metrics_ref,bredin17_interspeech,kuhn55_hungarian}. We apply a 0.5\,s forgiveness collar around reference boundaries (i.e., exclude a $\pm$0.25\,s window around each annotated speaker-change from scoring to tolerate boundary imprecision) \cite{pyannote_metrics_ref,bredin17_interspeech}. 


\textbf{Real-time factor (RTF).} To quantify throughput, we report RTF, defined as the ratio of total wall-clock processing time to the corresponding input audio duration \cite{heitkaemper24_interspeech}:
\begin{equation}
\mathrm{RTF} = \frac{t_{\mathrm{proc}}}{t_{\mathrm{audio}}},
\end{equation}
where $t_{\mathrm{proc}}$ is the end-to-end elapsed time required to process an utterance and $t_{\mathrm{audio}}$ is the duration of the input audio. RTF can be interpreted as the \emph{seconds of compute per second of audio}: where $\mathrm{RTF}<1$ indicates faster-than-real-time processing, $\mathrm{RTF}\approx 1$ indicates real-time operation, and $\mathrm{RTF}>1$ indicates the system falls behind live audio \cite{heitkaemper24_interspeech}.

\textbf{Latency.} We operate in chunk-wise streaming inference with bounded look-ahead and report \textbf{algorithmic latency} (i.e., delay induced by buffering future context), distinct from hardware-dependent computational delay \cite{rahou24_interspeech}. 
Let the \textbf{chunk length (update period)} be $C$ seconds and the \textbf{right-context (look-ahead)} be $R$ seconds. At each update, the system buffers an additional $R$ seconds of future audio and emits speaker labels in blocks of length $C$.

We summarize the latency as the effective decision delay induced by these two control parameters, which is typically modeled as the sum of chunk length and right-context length \cite{rahou24_interspeech,s2snd_arxiv}:
\[
\text{Algorithmic latency} \approx C + R.
\]

\textbf{Right-context sweep.}
We fix $C=1.0$~s and sweep $R$ over log-spaced values to quantify the effect of additional right-context.

\textbf{Chunk length sweep.}
We fix $R=0$ and sweep $C$ over log-spaced values to quantify the effect of the chunk length.

As linear spacing in seconds would overemphasize the high-latency settings, we use log-spaced sweeps because these latency control parameters are strictly positive and their impact is typically scale-sensitive per standard practice for tuning positive, scale-sensitive hyperparameters \cite{feurer_hyperparameter_2019,bergstra12a}.

\subsection{Pruning}
\textbf{Baseline.} In our experiments, we keep the embedding/clustering backend and released pipeline thresholds fixed (Sec.~\ref{sec:setup}), and apply all pruning (and any recovery fine-tuning) only to the segmentation model parameters.

We apply iterative, magnitude-based structured pruning to the SIMSAMU segmentation model\cite{hf_simsamu_segmentation,hf_pyannote_segmentation30}. Our goal is to perform gradual pruning with recovery fine-tuning to obtain structured throughput and memory savings while preserving accuracy \cite{han15_deepcompression,zhu17_prune,li17_pruningfilters}. We use three disjoint subsets to avoid evaluation leakage: \emph{train} for any recovery fine-tuning after pruning, \emph{calib} reserved for calibration-only procedures used later (quantization), and \emph{eval} for metric reporting. For each target pruning ratio $\rho \in \{0.2,0.4,0.6,0.8\}$, we reload the baseline model and prune gradually over $K$ iterations. In each iteration, we remove an additional fraction $\Delta\rho=\rho/K$ of the lowest-importance units, so the model reaches the same final pruning level with better stability than one-shot pruning \cite{han15_deepcompression,zhu17_prune}.

Let $u$ denote a prunable \emph{unit} in the segmentation network. Depending on the pruning structure, $u$ is either (i) a BiLSTM hidden unit (grouped across gates and directions) or (ii) a neuron/channel in the post-LSTM linear stack. Let $\theta_u$ be the set of parameters associated with $u$. We score units by weight magnitude,
\begin{equation}
s(u) = \lVert \theta_u \rVert_F,
\end{equation}
and at each step prune the $\Delta\rho$ fraction of remaining units with the smallest $s(u)$ \cite{han15_deepcompression,zhu17_prune}.

\textbf{Hidden (BiLSTM hidden size).}
We treat each BiLSTM hidden unit as a unit $u$ (grouped across the four LSTM gates and both directions). Concretely, for a hidden index $c$, $\theta_u$ collects the corresponding gate rows/blocks in the recurrent weights (and associated biases). We prune the lowest-scoring hidden units and rebuild the BiLSTM with a smaller hidden size.

\textbf{Linear (post-LSTM linear channels).}
We prune neurons/channels in the post-LSTM linear stack. For an intermediate channel $c$, we score it by the magnitude of its incident weights in adjacent linear layers (e.g., column in the incoming linear weight and row in the outgoing linear weight) and rebuild narrower linear layers by keeping the top-scoring channels.

After each pruning iteration, we run a short supervised fine-tuning phase on \emph{train} to recover performance, using RTTM-derived speech activity targets with \texttt{BCEWithLogitsLoss} and AdamW \cite{han25_interspeech,han15_deepcompression,zhu17_prune}. Final pruned variants are reported with DER, RTF, parameter count, and model size (Table~\ref{tab:pruning}). Parameters are reported as the total number of trainable weights and model size is the on-disk serialized checkpoint size.

\subsection{Quantization}
We evaluate six numeric-precision modes for the segmentation model (Table~\ref{tab:pruning}).
Quantization is applied to the \emph{selected pruned model by lowest DER}, and we ensure that evaluation clips never appear in training or calibration \cite{Jacob_2018_CVPR,pytorch_quantization_docs}. \textit{We denote compute precision as W$b_w$A$b_a$, where $b_w$ and $b_a$ are the weight and activation bit-widths, respectively.} For the INT4 setting, we use W4A8 as activations are typically more distribution-shifted and outlier-prone \cite{xiao23_smoothquant,dettmers22_llmint8}. Keeping them at 8-bit improves stability. FP32, floating-point baselines, runs the released pipeline reference without modification. FP16 uses mixed-precision inference with CUDA autocast \cite{pytorch_amp_docs}. All quantized variants described below are implemented with PyTorch AO (\texttt{torchao}). For post-training quantization (PTQ), we replace each \texttt{nn.Linear} layer in the segmentation model with a quantized inference module and calibrate activation scales on \emph{calib} only, following standard static PTQ practice \cite{Jacob_2018_CVPR,pytorch_ptq_tutorial}. We report INT8\_PTQ and INT4\_PTQ. Low-bit PTQ can be sensitive to activation and weight outliers, so we follow the standard calibration procedure on \emph{calib} to set quantization ranges \cite{Jacob_2018_CVPR,pytorch_ptq_tutorial}. For quantization-aware training (QAT), we replace \texttt{nn.Linear} layers with quantized training modules, fine-tune on \emph{train} split, then convert to real quantized inference operators \cite{Jacob_2018_CVPR,pytorch_quantization_docs,torchao_qat_docs}. After conversion, we run activation calibration on \emph{calib} and evaluate on \emph{eval}.

For each model, we report DER, RTF, and model size, while parameter count remains unchanged. We also report the per-recording change in DER ($\Delta$DER) relative to the FP32 baseline, summarized by the median, interquartile range (IQR), and tail statistics (90th percentile and maximum).

\section{Results}

\subsection{Latency: Does additional delay change diarization accuracy?}

\textit{With a fixed chunk length ($C=1.0$~s), does increasing right-context ($R$) change performance?}
\begin{figure}[t]
  \centering
  \includegraphics[width=0.90\linewidth]{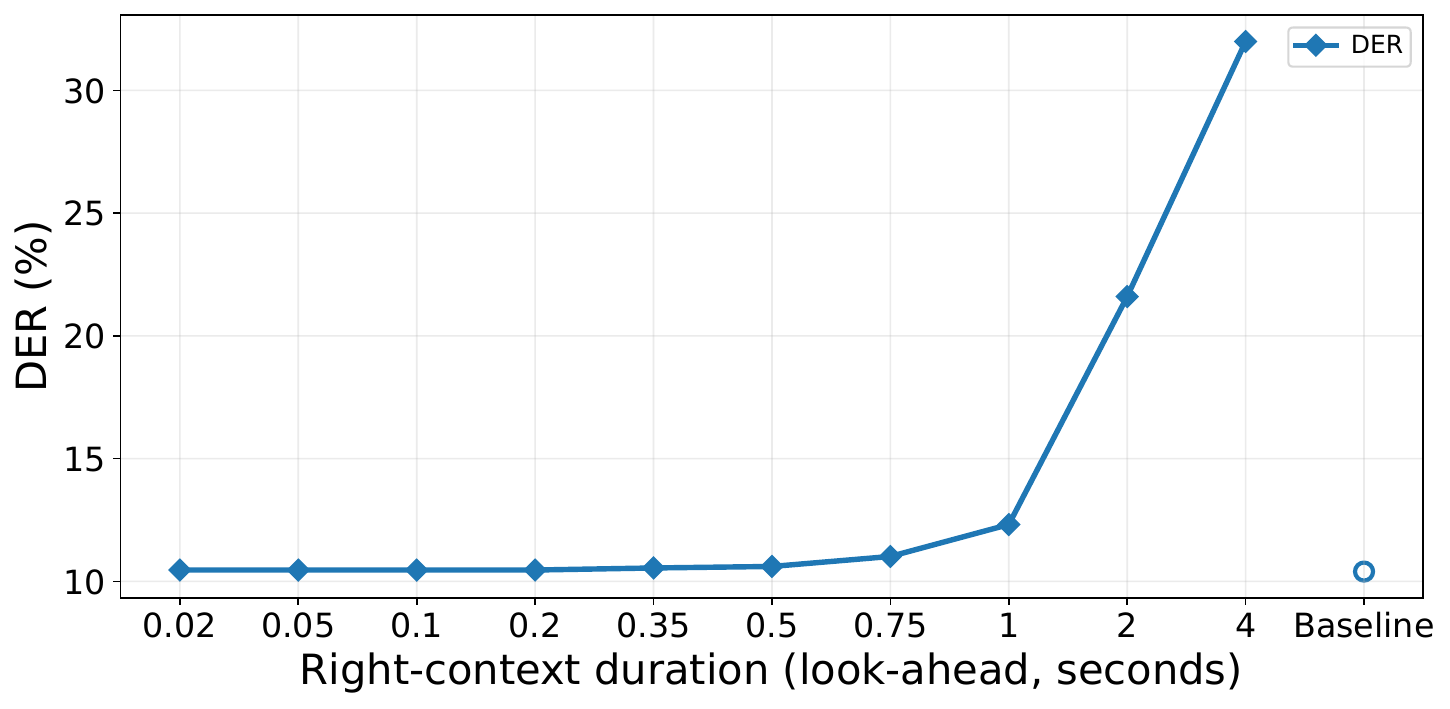}
  \caption{Effect of right-context $R$ (seconds, $x$-axis) on DER (\%, $y$-axis ) at fixed chunk length. DER is mostly stable for small $R$ and increases sharply at larger right-context values.}
  \label{fig:right_context}
\end{figure}
Fig.~\ref{fig:right_context} shows that performance is largely stable for small-to-moderate $R$. Large right-context values can shift performance noticeably, indicating that increasing $R$ is not consistently beneficial in this pipeline. The sharp change around $R\!\approx\!1$\,s is consistent with the typical turn-changes in dispatch calls, where buffering that much future audio can blur speaker-change boundaries.

\textit{With zero right-context ($R=0$), how does the chunk length ($C$) change performance?}
\begin{figure}[t]
  \centering
  \includegraphics[width=0.90\linewidth]{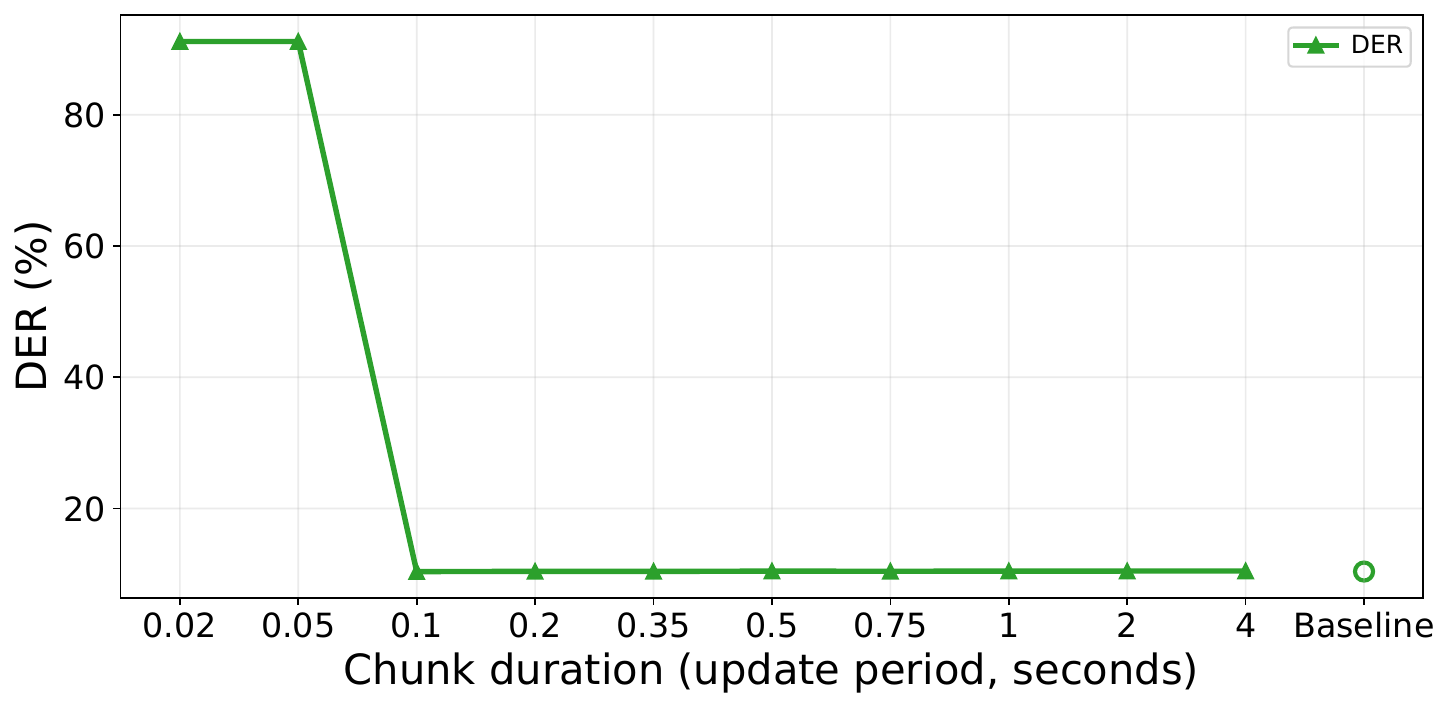}
  \caption{Effect of chunk length duration $C$ (seconds, $x$-axis) on DER (\%, $y$-axis) with right-context fixed at $R=0$. Very small chunks degrade DER sharply, while moderate-to-large chunks reaches a stable plateau.}
  \label{fig:chunk_sweep}
\end{figure}
Fig.~\ref{fig:chunk_sweep} shows that very small chunks (0.02--0.10\,s) substantially change error, while $C\gtrsim 0.1$~s yields a stable plateau where further increases in $C$ cause only small changes. This suggests that moderate chunk lengths are sufficient for stable behavior, with diminishing returns if the length is increased further.

\subsection{How do structured pruning level and type impact accuracy and efficiency?}

\begin{table}[t]
\caption {Hidden-unit pruning yields large model-size reductions but degrades accuracy sharply, while linear-channel pruning preserves accuracy better at low-to-moderate levels with only modest model size changes. Baseline denotes the released unpruned FP32 segmentation model.} 
\footnote {Linear--40: Linear pruning with 40\% pruned} 
\centering
\scriptsize
\setlength{\tabcolsep}{1.0pt}
\renewcommand{\arraystretch}{1.00}
\resizebox{\columnwidth}{!}{%
\begin{tabular}{%
C{1.75cm}  
C{1.05cm}  
C{1.05cm}  
C{0.80cm}  
C{0.85cm}  
}
\toprule
Variant & Params (M) & Size (MB) & RTF & DER (\%) \\
\midrule
Baseline         & 1.47 & 5.63 & 0.0232 & 10.70 \\
\midrule
Hidden--20       & 1.01 & 3.86 & 0.0225 & 19.24 \\
Hidden--40       & 0.69 & 2.65 & 0.0221 & 27.71 \\
Hidden--60       & 0.47 & 1.82 & 0.0224 & 35.88 \\
Hidden--80       & 0.33 & 1.26 & 0.0224 & 50.47 \\
\midrule
Linear--20       & 1.46 & 5.59 & 0.0225 & 14.94 \\
\rowcolor{green!15}
Linear--40       & 1.45 & 5.55 & 0.0223 & 12.66 \\
Linear--60       & 1.45 & 5.52 & 0.0222 & 25.97 \\
Linear--80       & 1.44 & 5.50 & 0.0222 & 35.86 \\
\bottomrule
\end{tabular}%
}
\label{tab:pruning}
\end{table}

Table~\ref{tab:pruning} shows a clear accuracy--compression trade-off, but the effect depends strongly on the pruning structure. Hidden-unit pruning substantially reduces parameters and model size, yet accuracy degrades sharply even at low pruning levels and worsens rapidly. In contrast, linear-channel pruning changes model size only marginally and preserves accuracy better at low-to-moderate levels, with Linear--40 achieving the lowest DER among pruned models. Across all variants, RTF remains nearly unchanged ($\approx$0.022--0.023), indicating that structured pruning provides limited end-to-end RTF gains in this pipeline even when parameter counts drop. This likely occurs because the measured end-to-end runtime is dominated by non-pruned pipeline stages and kernel overheads, so reducing segmentation parameters does not translate into proportional RTF speedups.


\subsection{How do quantization modes impact model size and accuracy under model pruning?}

\begin{table}[t]
\caption{Quantization sweep on the pruned segmentation model (Linear--40). Lower precision substantially reduces model size, but end-to-end RTF changes are small under this pipeline. FP16 yields the lowest DER among quantized models.}
\label{tab:quant}
\centering
\footnotesize
\setlength{\tabcolsep}{3.8pt}
\renewcommand{\arraystretch}{1.05}
\resizebox{\columnwidth}{!}{%
\begin{tabular}{c c c c c}
\toprule
Mode & Precision (W/A) & Size (MB) & RTF & DER (\%) \\
\midrule
FP32      & W32A32 & 5.55 & 0.0223 & 12.66 \\
\rowcolor{green!15}
FP16      & W16A16 & 2.78 & 0.0250 & 15.02
\\
INT8\_PTQ & W8A8   & 1.54 & 0.0227 & 23.67 \\
INT8\_QAT & W8A8   & 1.54 & 0.0226 & 19.02 \\
INT4\_PTQ & W4A8   & 0.89 & 0.0227 & 47.80 \\
INT4\_QAT & W4A8   & 0.89 & 0.0229 & 44.93 \\
\bottomrule
\end{tabular}%
}
\end{table}

\begin{figure}[t]
  \centering
  \includegraphics[width=0.98\columnwidth]{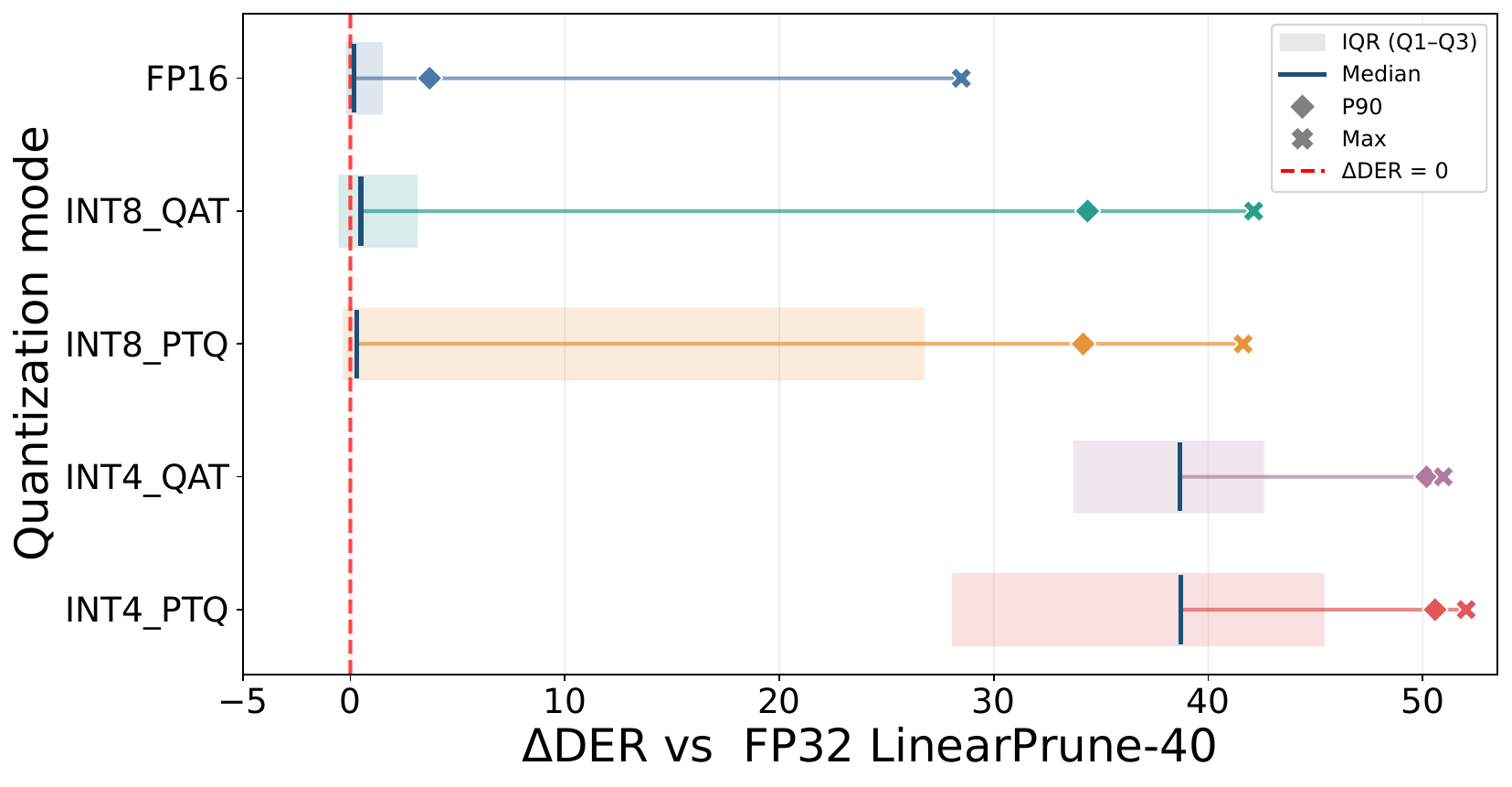}
  \caption{Per-recording diarization error change under quantization on Linear--40
  The x-axis shows $\Delta$DER (percentage points) and the y-axis lists quantization modes. The red dashed line at $\Delta\mathrm{DER}=0$ indicates parity with the FP32 Linear--40 baseline. Diamonds mark the 90th percentile (P90) and $\times$ marks the maximum $\Delta$DER across recordings. FP16 is closest to baseline, with the smallest median $\Delta$DER and a tight IQR.}
  \label{fig:quant_delta_der}
\end{figure}


Table~\ref{tab:quant} summarizes the quantization trade-offs on the selected pruned model (Linear--40). Among lower-precision variants, \textbf{FP16} attains the lowest DER, outperforming both INT8 modes, while INT4 variants substantially degrade accuracy. Across both INT8 and INT4, QAT improves over PTQ since fine-tuning lets the model adapt to quantization noise and reduces rounding error at inference.
Relative to the unpruned FP32 baseline, the final Linear--40 + FP16 variant is $2.0\times$ smaller at a 40\% relative DER increase, while RTF is essentially unchanged.

To assess stability, Fig.~\ref{fig:quant_delta_der} reports the distribution of per-recording $\Delta$DER relative to FP32 Linear--40. FP16 is closest to baseline, with a small median increase and a tight interquartile range. INT8 variants exhibit modest median degradation but substantially higher variability, including high-$\Delta$DER outliers. INT8\_QAT reduces this spread compared to INT8\_PTQ. In contrast, INT4 (PTQ/QAT) yields a consistently large positive shift across the distribution, indicating systematic accuracy loss.


\section{Discussion}
This work is a first step toward making \emph{efficiency claims in diarization} more interpretable and comparable across studies. Many diarization pipelines are evaluated with papers reporting DER, but omitting the key deployment settings (e.g., chunk length, look-ahead, precision). Our results show that these configuration choices can influence accuracy, so two systems may not be directly comparable even when they use the same high-level architecture. While this study does not introduce a new diarization model, it characterizes settings that drive the accuracy--efficiency trade-off and calls for attention to report them for resource-constrained inference.

A key implication is that model compression does not automatically translate to end-to-end throughput gains in a complete diarization stack. While low-bit quantization and structured pruning can reduce model storage, the measured RTF changes only marginally, suggesting that runtime may be dominated by non-quantized components (feature extraction, I/O, clustering) or kernel effects. For deployment-oriented speech research, this motivates evaluating efficiency at the \emph{pipeline level} rather than at checkpoints with faster diarization. 

\textbf{Practical considerations.}
These findings suggest several reporting practices for deployment-oriented diarization. Compression should be evaluated at the full-pipeline level, since reducing the segmentation model size may not improve end-to-end throughput. DER should remain the main operational metric, but segmentation-level diagnostics can help explain where errors enter the pipeline. Chunk-wise inference should also be distinguished from fully online diarization, which requires persistent speaker-state tracking. Accordingly, conclusions about accuracy--efficiency trade-offs should be tied to the specific pipeline, hardware setting, and quantization backend used.

\textbf{Limitations.}
SIMSAMU is a small, specialized medical-dispatch simulation corpus with two primary roles, so absolute DER levels and the observed compression thresholds may not generalize \cite{nun25_simsamu}. Our conclusions are based on a specific segmentation--embedding--clustering pipeline; different architectures could change sensitivity to latency settings (right-context and chunking) and to pruning or reduced-precision arithmetic. Finally, efficiency is measured in one runtime environment, and throughput-memory observations can vary with kernels and hardware generation.

\textbf{Future work.}
Future work can evaluate additional languages and more diverse interaction conditions (e.g., third-party speakers, background speech, and higher overlap), and jointly optimize latency parameters ($C$ and $R$) with pruning and quantization to map the full accuracy--efficiency trade-off. It can also isolate contributions from the segmentation versus embedding and clustering stages, and include robustness analysis under higher overlap to quantify degradation under compression.

\section{Acknowledgments}

We thank Colby HPC Cluster and especially Stephen Cousins for their support with the compute resources.

\section{Generative AI Use Disclosure}
Generative AI tool (ChatGPT) was used for LaTex table formatting assistance.

\bibliographystyle{IEEEtran}
\bibliography{references}

\end{document}